\documentclass[12pt]{article}
\usepackage[english]{babel}
\usepackage{amsmath}
\usepackage{graphicx,psfrag,epsf}
\usepackage{enumerate}
\usepackage{natbib}
\usepackage{amsfonts}
\usepackage[latin1]{inputenc}
\usepackage[ulem=normalem]{changes}
\usepackage{cancel}
\usepackage{paralist}
\usepackage{multirow}
\usepackage[small]{caption}
\usepackage{setspace}
\usepackage{url}
\usepackage{bbm}
\usepackage{dsfont}
\usepackage{epstopdf}

\newcommand{\blind}{0}

\begin{document}

\def\spacingset#1{\renewcommand{\baselinestretch}%
{#1}\small\normalsize} \spacingset{1}

\if0\blind
{
  \title{\bf Spatial Depth-Based Classification for Functional Data}
  \author{
    \textbf{Carlo Sguera}\\
    \normalsize Department of Statistics, Universidad Carlos III de Madrid\\
    \normalsize 28903 Getafe (Madrid), Spain\\
    \normalsize(\texttt{csguera@est-econ.uc3m.es})\\
   \textbf{Pedro Galeano} \\
    \normalsize Department of Statistics, Universidad Carlos III de Madrid\\
    \normalsize 28903 Getafe (Madrid), Spain\\
    \normalsize(\texttt{pedro.galeano@uc3m.es})\\
     and \\
    \textbf{Rosa E. Lillo} \\
    \normalsize Department of Statistics, Universidad Carlos III de Madrid\\
    \normalsize 28903 Getafe (Madrid), Spain\\
    \normalsize(\texttt{rosaelvira.lillo@uc3m.es})\\
}
    \date{}
  \maketitle
} \fi

\if1\blind
{
  \bigskip
  \bigskip
  \bigskip
  \begin{center}
    {\LARGE\bf Spatial Depth-Based Classification for Functional Data}
\end{center}
  \medskip
} \fi

\begin{abstract}
We enlarge the number of available functional depths by introducing the kernelized functional spatial depth (KFSD). KFSD is a local-oriented and kernel-based version of the recently proposed functional spatial depth (FSD) that may be useful for studying functional samples that require an analysis at a local level. In addition, we consider supervised functional classification problems, focusing on cases in which the differences between groups are not extremely clear-cut or the data may contain outlying curves. We perform classification by means of some available robust methods that involve the use of a given functional depth, including FSD and KFSD, among others. We use the functional \textit{k}-nearest neighbor classifier as a benchmark procedure. The results of a simulation study indicate
that the KFSD-based classification approach leads to good results. Finally, we consider two real classification problems, obtaining results that are consistent with the findings observed with simulated curves.
\end{abstract}

\noindent%
\vfill
\hfill


\newpage
\spacingset{1.45} 

\section{INTRODUCTION}
\label{sec:intro}

The technological advances of the last decades in fields such
as chemometrics, engineering, finance, growth analysis or
medicine have allowed to observe random samples of
curves. In these cases, it is common to assume that the sample has been
generated by  a stochastic function, namely a random variable
taking values on an infinite-dimensional space. To analyze this
type of data, it is convenient to use the tools provided by a
recent area of statistics known as functional data analysis (FDA).
For two complementary FDA overviews, one parametric and the other
nonparametric, see \cite{RamSil2005} and \cite{FerVie2006},
respectively, whereas for inference and asymptotic theory for functional data see \cite{HorKok2012}. 

When data are curves, there are at least three reasons
why FDA should be preferred to a standard multivariate data analysis. First, although curves are usually observed as vectors,
the evaluation points may differ in number and/or position from
curve to curve and, unlike multivariate observations, they cannot
be permuted. Second, any  stochastic function has a dependence
structure and, consequently, functional data are usually rather
autocorrelated and hard to be analyzed with standard multivariate
procedures. Third, functional samples may contain less curves than
evaluation points, and great difficulties arise when this feature
occurs in multivariate data analysis.

Despite the previous reasoning, many multivariate techniques have
inspired advances in FDA, and a good example is the notion of multivariate depth. According to
\cite{Ser2006}, a multivariate depth is a function that provides
a $P$-based center-outward ordering of points in $\mathbb{R}^{d}$, where $P$ is a probability distribution on
$\mathbb{R}^{d}$. Hence, the values of any depth measure should
be higher at points that are central relative to the
probability distribution $P$, and lower at peripheral points for $P$. For an overview on multivariate depths, see for example \cite{ZuoSer2000}. An implementation of the notion of multivariate depth is the spatial depth (SD, \citeauthor{Ser2002} \citeyear{Ser2002}), which is defined as follows: let
$\mathbf{Y}$ be a $d$-dimensional random vector having cumulative distribution
function $F$. Then,  the multivariate spatial depth of $\mathbf{x} \in \mathbb{R}^{d}$
relative to $F$ is defined as

\begin{equation}
\label{eq:mulDep} SD(\mathbf{x}, F) = 1 - \left\|\int
S(\mathbf{x}-\mathbf{y})\, dF(\mathbf{y})\right\|_{E} = 1 -
\left\|\mathbb{E}\left[
S(\mathbf{x}-\mathbf{Y})\right]\right\|_{E},
\end{equation}

\noindent where $\|\mathbf{\cdot}\|_{E}$ is the Euclidean norm in
$\mathbb{R}^{d}$ and  $S: \mathbb{R}^d \rightarrow \mathbb{R}^d$
is the multivariate spatial sign function given by
\begin{equation}
\label{eq:mulSign} S(\mathbf{x}) = \left\{
\begin{array}{cr}
\frac{\mathbf{x}}{\|\mathbf{x}\|_{E}}, & \mathbf{x} \neq \mathbf{0}, \\
\mathbf{0}, & \mathbf{x} = \mathbf{0}.
\end{array}
\right.
\end{equation}

The spatial depth has a connection with the notion of spatial quantile introduced by \cite{Cha1996}. Let
$\{\mathbf{u} \colon \mathbf{u} \in \mathbb{R}^{d} ,
\|\mathbf{u}\|_{E} < 1\}$, then $Q_{F}(\mathbf{u})$ is the
$\mathbf{u}$th spatial quantile of  $\mathbf{Y}$  if and only if
$Q_{F}(\mathbf{u})$ is the value of $\mathbf{q}$ which minimizes

\begin{equation*}
\label{eq:mulQua01} \mathbb{E}\left[\Phi(\mathbf{u},
\mathbf{Y}-\mathbf{q}) - \Phi(\mathbf{u}, \mathbf{Y})\right],
\end{equation*}

\noindent where, for $\mathbf{y} \in \mathbb{R}^{d}$, $\Phi(\mathbf{u}, \mathbf{y})=\|\mathbf{y}\|_{E} + \langle
\mathbf{u},\mathbf{y}\rangle_{E}$, and $\langle \mathbf{u},
\mathbf{y}\rangle_{E}$ is the Euclidean inner product of
$\mathbf{u}$ and $\mathbf{y}$. Under some mild conditions, $Q_{F}(\mathbf{u})$ and $SD(\mathbf{x},F)$ are linked in the following way:

\begin{equation*}
\label{eq:mulCon} \|Q_{F}^{-1}(\mathbf{x})\|_{E} =1-SD(\mathbf{x}, F).
\end{equation*}

A key point for our work is that both $SD(\mathbf{x}, F)$ and $Q_{F}(\mathbf{u})$ extend naturally from $\mathbb{R}^{d}$ to any infinite-dimensional Hilbert space $\mathbb{H}$. First, \cite{ChaCha2013} defined a functional version of $SD(\mathbf{x}, F)$, the functional spatial depth function $FSD(x, P)$, where $x \in \mathbb{H}$ and $P$ is a probability distribution on $\mathbb{H}$. Second, \cite{Cha1996} defined a functional version of $Q_{F}(\mathbf{u})$, the functional spatial quantile function $FQ_{P}(u)$, where $\left\{u:u \in \mathbb{H}, \|u\| < 1\right\}$ and $\|\cdot\|$ is the norm derived from the inner product $\langle\cdot,\cdot\rangle$ in $\mathbb{H}$. The first contribution of this paper is to show that $FSD(x, P)$ and $FQ_{P}(u)$ are linked, as well as $SD(\mathbf{x}, F)$ and $Q_{F}(\mathbf{u})$.

It is worth noting that $SD(\mathbf{x}, F)$ depends on the whole $F$. In other words, the multivariate spatial depth tackles data ordering through a global approach, as well as other well-known multivariate depths do, e.g., the halfspace depth (\citeauthor{Tuk1975} \citeyear{Tuk1975}) or the simplicial depth (\citeauthor{Liu1990} \citeyear{Liu1990}). However, in some cases it may be useful to study more in detail narrower neighborhoods of the observations, and therefore to dispose of depth functions coherent with a local approach. To this end, there are multivariate references focused on the idea of introducing a local depth. For example, \cite{AgoRom2011} proposed the local simplicial depth, which considers only random simplices with size not greater than a fixed threshold, and the local halfspace depth, which replaces halfspaces with closed slabs, whereas \citet*{CheDanPenBar2009} proposed the kernelized spatial depth (KSD), which is a local-oriented and kernel-based version of the multivariate spatial depth SD. Moving to the functional framework, and since $FSD(x, P)$ is also global-oriented as it depends on the whole $P$, the second contribution of this paper consists in proposing the kernelized functional spatial depth (KFSD), that is a local-oriented and kernel-based version of FSD.

Besides FSD and KFSD, other implementations of the idea
of functional depth already exist in the FDA literature: \cite{FraMun2001} defined the Fraiman and Muniz depth (FMD), which
tries to measure how long remains a curve in the middle of a sample. \citet{CueFebFra2006} proposed the h-modal depth
(HMD), which tries to measure how densely a curve is surrounded by
other curves. \cite{CueNie2008} and \cite{CueFra2009} proposed the random Tukey depth (RTD) and the integrated dual
depth (IDD), respectively. Both RTD and IDD are based on the computation of $p$ random
one-dimensional projections of the curves, but they differ in the way the $p$-dimensional vectors of projections are managed:
RTD makes use of the halfspace depth, whereas IDD of the simplicial depth. Finally, \cite{LopRom2009} proposed
the band and modified band depths (BD and MBD, respectively). Their second proposal MBD is based on all the possible bands defined by
the graphs on the plane of $2, 3, \ldots$ and $J$ curves, and on a
measure of the sets where another curve is inside these bands.

Functional depths are
useful to perform exploratory FDA and to build robust functional
methods. In particular, in this paper we tackle the supervised
functional classification problem
by considering three depth-based procedures: the
distance to the trimmed mean method (DTM, \citeauthor*{LopRom2006}
\citeyear{LopRom2006}), the weighted averaged distance method
(WAD, \citeauthor*{LopRom2006} \citeyear{LopRom2006}) and the
within maximum depth method (WMD, \citeauthor*{CueFebFra2007}
\citeyear{CueFebFra2007}). The depth-based classification
methods have been mainly proposed for functional datasets that are
possibly affected by the presence of outlying curves and actually,
since the available FDA outlier detection procedures are still few
(see for example \citeauthor*{FebGalGon2008}
\citeyear{FebGalGon2008}), robustness may be a key issue in many
functional classification problems. For this reason, the
depth-based methods try to classify curves in a robust way: DTM and WAD use the depth information provided by the
curves of a training sample, whereas WMD looks at the within group depth values of
the curves to classify. The third contribution of this work consists in analyzing the performances of these depth-based procedures when used
together with the above-mentioned existing depth functions and with our proposal, the kernelized functional spatial depth KFSD.

The remainder of the article is organized as follows. In Section
\ref{sec:FSDs} we recall the definition of FSD and introduce KFSD. In Section \ref{sec:classification} we consider the supervised functional
classification problem and show the details of the DTM, WAD
and WMD methods. The results of a simulation study and two real data analyses are presented in Sections \ref{sec:simStudy} and
\ref{sec:real}, respectively. Both the simulation study and the real data studies show
that a KFSD-based approach generally leads to good
results, especially when WMD and KFSD are combined together. Some conclusions are drawn in
Section \ref{sec:conc}.

\section{FUNCTIONAL SPATIAL DEPTHS}
\label{sec:FSDs}

A random variable $Y$ is called functional
random variable if it takes values in a functional space. A functional
dataset consists in the observations of $n$ functional random
variables identically distributed as $Y$ (\citeauthor*{FerVie2006}
\citeyear{FerVie2006}). In what follows, we assume that the
functional space is a Hilbert space, denoted by $\mathbb{H}$, with
norm $\|\cdot\|$ inherited from the inner product $\langle \cdot,
\cdot\rangle$ in $\mathbb{H}$.

In this section, we introduce the functional spatial depth (FSD), recently proposed by \cite{ChaCha2013}, and the kernelized functional spatial depth (KFSD), which is a new depth measure for functional data. Both FSD and KFSD rely on the general idea of spatial depth. The origins of the spatial approach date back to \cite{Bro1983}, who studied the problem of robust location estimation for two-dimensional spatial data and introduced the idea of spatial median. This approach considers the geometry of the data and is the basis for the notions of multivariate spatial depth function (\citeauthor*{Ser2002} \citeyear{Ser2002}) and multivariate spatial quantiles (\citeauthor*{Cha1996} \citeyear{Cha1996}), both already introduced in Section \ref{sec:intro}. Note that (\ref{eq:mulDep}) and (\ref{eq:mulSign}) are practically two particular cases of two more general definitions for elements belonging to normed vector spaces. In this paper, we consider two different applications of (\ref{eq:mulDep}) and (\ref{eq:mulSign}): first, the spatial sign function for $x \in \mathbb{H}$, which is given by

\begin{equation*}
\label{eq:funSign}
FS(x) = \left\{
\begin{array}{cr}
\frac{x}{\|x\|}, & x \neq 0, \\
0, & x = 0.
\end{array}
\right.
\end{equation*}

\noindent Second, the spatial depth function for $x \in \mathbb{H}$, which is given by

\begin{equation*}
\label{eq:funDep}
FSD(x, P) = 1 - \left\|\mathbb{E}\left[FS(x-Y)\right]\right\|,
\end{equation*}

\noindent where $Y$ is a functional random variable with probability distribution $P$ on $\mathbb{H}$ (\citeauthor{ChaCha2013} \citeyear{ChaCha2013}). 

When a sample of curves is observed, say $(y_{i})_{i=1,\ldots, n}$, $FSD(x, P)$ must be replaced with its corresponding sample version, i.e.,

\begin{equation}
\label{eq:funFSDn}
FSD_{n}(x) = 1 - \frac{1}{n} \left\|\sum_{y \in (y_{i})_{i=1, \ldots, n}} FS(x-y)\right\|.
\end{equation}

\noindent Since in practice $x$ and $(y_{i})_{i=1,\ldots, n}$ are observed at discretized and finite sets of domain points, and since these sets may differ from one curve to another and/or may not contain equidistant points, the computation of $FSD_{n}(x)$ may require the estimation of $x$ and $(y_{i})_{i=1,\ldots, n}$ at a common set of equidistant domain points.

As in $\mathbb{R}^d$, the notion of functional spatial depth can be related to the notion of functional spatial quantiles. Let $\{u \colon u \in \mathbb{H} , \|u\| < 1\}$, then the $u$th functional spatial quantile of the $\mathbb{H}$-valued random variable $Y$ with probability distribution $P$ is obtained by minimizing

\begin{equation}
\label{eq:funQua01}
\mathbb{E}\left[\Phi(u, Y-q) - \Phi(u, Y)\right],
\end{equation}

\noindent with respect to $q$, where, for $y \in \mathbb{H}$, $\Phi(u, y)=\|y\|+\langle u, y\rangle$, and $\langle u, y\rangle$ is the inner product of $u$ and $y$ (\citeauthor{Cha1996} \citeyear{Cha1996}). If $Y$ is not concentrated on a straight line and is not strongly concentrated around single points, \citet*{CarCenZit2011} showed that the Fr\'echet derivative of the convex function in (\ref{eq:funQua01}) is given by

\begin{equation}
\label{eq:funQua02}
\Phi(q)=-\mathbb{E}\left[\frac{Y-q}{\|Y-q\|}\right] - u,
\end{equation}

\noindent and that the $u$th quantile of $Y$ is given by the unique solution of the equation $\Phi(q)=0$. If $x$ is the solution of $\Phi(q)=0$, the previous results provide a functional spatial quantile function, say $FQ_{P}$. Then, considering the norm of its inverse evaluated at $x$, we have that

\begin{equation*}
\label{eq:funCon}
\|FQ_{P}^{-1}(x)\| = \left\|-\mathbb{E}\left[\frac{Y-x}{\|Y-x\|}\right]\right\| = \left\|\mathbb{E}\left[FS(x-Y)\right]\right\| = 1 - FSD(x, P),
\end{equation*}

\noindent which shows that the direct connection between the notions of spatial depth and quantiles holds also in functional Hilbert spaces, and it enriches FSD with an interesting interpretability property.

\cite{ChaCha2013} showed some other properties of FSD: (1) $FSD(x,P)$ is invariant under the class of linear transformations $T:\mathbb{H} \rightarrow \mathbb{H}$, where $T(x)=cAx+b$ for some $c>0, b \in \mathbb{H}$ and an isometry $A$ on $\mathbb{H}$; (2) if $P$ is non-atomic, then $FSD(x,P)$ is continuous in $x$; (3) if $\mathbb{H}$ is strictly convex and $P$ is non-atomic and not supported on a line in $\mathbb{H}$, then $FSD(x,P)$ has a unique maximum at the spatial median $m$ of $Y$\footnote{If $Y$ is not concentrated on a straight line and is not strongly concentrated around single points, the spatial median $m$ of $Y$ is the unique solution of (\ref{eq:funQua02}) for $u$ equal to the zero element in $\mathbb{H}$.} and its maximum value is 1; (4) for any fixed non-zero $x \in \mathbb{H}$ and sequence $\left\{m+nx\right\}_{n \in \mathbb{N^{+}}}$, the following holds: $FSD(m+nx,P) \rightarrow 0$ as $n \rightarrow \infty$; (5) $FSD(x,P)$ does not suffer from degeneracy for many infinite dimensional probabilities distributions (for more details on property (5), see \citeauthor{ChaCha2013} \citeyear{ChaCha2013}).  

We would like to emphasize that, for any $x$ and $y \in (y_{i})_{i=1, \ldots, n}$, the functional spatial sign
function $FS(x-y)$ is a unit-norm curve that can be interpreted as
the direction from $x$ to $y$. Therefore, $FSD_{n}(x)$ depends on
the sum of $n$ directions that contributes equally to
$FSD_{n}(x)$. This feature is a key property of $FSD_{n}(x)$, but
it generates a trade-off: on one side, it makes $FSD_{n}(x)$
robust to the presence of outliers; on the other, it transforms
$(x-y)$ into a unit-norm curve regardless of $y$ being a
neighboring or a distant curve from $x$. As observed by
\cite{CheDanPenBar2009} in the multivariate framework, in some
circumstances a more local analysis of the curves may be of
interest and it would allow the information brought by $y$
to depend on the value of a certain distance between $x$ and $y$. 

\begin{figure}[!htbp]
\centering
\includegraphics[scale=0.5]{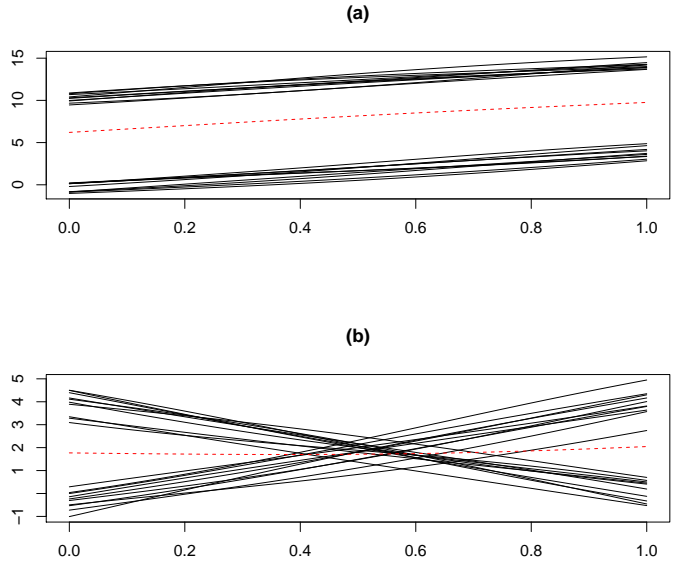}
\captionsetup{width=0.8\textwidth}
\caption{Two functional datasets that we use to show the differences between global and local depths.}
\label{fig:globVSloc}
\end{figure}

To illustrate the differences between a global and a local functional depth approach, let us consider two center-outward ordering problems. In the first example, we generate $21$ curves from a given process and divide them in three groups with $10$, $10$ and 1 curves, respectively. Then, we add a different constant to each group ($0$, $10$ and $5$, respectively), obtaining the curves at the top of Figure \ref{fig:globVSloc}. Afterwards, we compute the depth values of all the curves using the global-oriented depths FMD, RTD, IDD, MBD and FSD, and observe that the curve of the third group attains the highest depth with all the global depths. 

In the second example, the structure of the dataset is similar, but we use different transformations to obtain the curves belonging to the second and third group (see the plot at the bottom of Figure \ref{fig:globVSloc}). Also in this case, we observe that, according to the global depths, the curve of the third group turns out to be the deepest one. Clearly, both examples involve three strongly different classes of curves but, if we treat the curves of each example as belonging to a homogeneous sample, we observe rather inconvenient behaviors of the global depth functions. In the two examples, the curve in the third group is roughly in the geometric center of the dataset. However, it is far from the remaining curves. Therefore, it would be more reasonable to observe a low depth value at this curve. Actually, this happens with the local-oriented depths HMD and KFSD (formally presented below), mainly because they reduce the contribution of distant curves to the depth value of a fixed curve. Indeed, HMD and KFSD assign the lowest depth value to the curves belonging to the third group in both examples, and this is due to their local approach.

A common way to implement a local approach is to consider
kernel-based methods. To achieve this, and based on \citet{CheDanPenBar2009}, our first step consists in recoding the data. More in detail, instead of considering $x \in \mathbb{H}$,
we consider $\phi(x) \in \mathbb{F}$, where $\phi: \mathbb{H} \to \mathbb{F}$ is an embedding map and $\mathbb{F}$ is a feature space. Then, we define the kernelized functional spatial depth as

\begin{equation}
\label{eq:S02_19}
KFSD(x, P) = 1 - \left\|\mathbb{E}\left[FS(\phi(x)-\phi(Y))\right]\right\| = FSD(\phi(x), P_{\phi}),
\end{equation}

\noindent where $\phi(Y)$ is a recoded version of $Y$ and $P_\phi$ is a recoded version $P$. In particular, $\phi$ can be defined implicitly by a positive definite kernel, $\kappa: \mathbb{H} \times \mathbb{H} \rightarrow \mathbb{R}$, through

\begin{equation}
\label{eq:innToKer_new}
\kappa(x,y) = \langle\phi(x),\phi(y)\rangle.
\end{equation}

\noindent For this reason, in practice, it is only required to define $\kappa$, and not $\phi$. 

Next, we introduce the sample KFSD. To do this, note that it is possible to write

\begin{gather}
\left\Vert \sum_{y\in (y_{i})_{i=1,\ldots ,n}}FS(x-y)\right\Vert ^{2}=
\notag \\
=\sum_{y,z\in (y_{i})_{i=1,\ldots ,n}}\frac{\langle x,x\rangle +\langle
y,z\rangle -\langle x,y\rangle -\langle x,z\rangle }{\sqrt{\langle
x,x\rangle +\langle y,y\rangle -2\langle x,y\rangle }\sqrt{\langle
x,x\rangle +\langle z,z\rangle -2\langle x,z\rangle }}. \label{eq:gloToLoc}
\end{gather}

\noindent Therefore, $FSD_{n}(x)$ in (\ref{eq:funFSDn}) can be written in terms of the usual inner product in $\mathbb{H}$. Both inner products and kernel functions can be seen as similarity measures, but kernels are more powerful than inner products. We propose to replace the inner product function with a positive definite and stationary kernel function, and then to pass from $FSD_{n}(x)$ to our definition of sample kernelized functional spatial depth as follows:

\begin{gather}
KFSD_{n}(x) = 1 - \nonumber \\
\frac{1}{n} \left(\sum_{\substack{y,z \in (y_{i})_{i=1, \ldots, n}; \\ y \neq x; z \neq x}}\frac{\kappa( x,x)+\kappa( y,z)-\kappa( x,y)-\kappa( x,z)}{\sqrt{\kappa( x,x)+\kappa( y,y)-2\kappa( x,y)}\sqrt{\kappa( x,x)+\kappa( z,z)-2\kappa( x,z)}}\right)^{1/2}.\label{eq:funKerDep}
\end{gather}

\noindent It is worth noting that, as at the population level (see (\ref{eq:S02_19})), KFSD and FSD are related also at the sample level: thanks to (\ref{eq:innToKer_new}), we substitute $\kappa(\cdot,\cdot)$ with $\langle\phi(\cdot),\phi(\cdot)\rangle$ in (\ref{eq:funKerDep}), and, using (\ref{eq:gloToLoc}), we get that 

\begin{equation*}
\label{eq:kerDepToDep}
KFSD_{n}(x) = 1 - \frac{1}{n} \left\|\sum_{\phi(y) \in (\phi(y_{i}))_{i=1, \ldots, n}} FS(\phi(x)-\phi(y))\right\| = FSD_{n}(\phi(x)),
\end{equation*}

\noindent where $FSD_{n}(\phi(x))$ is the functional spatial depth of $\phi(x)$ relative to the recoded sample $\phi((y_{i}))_{i=1,\ldots, n}$. Therefore, $KFSD(x,P)$ and $KFSD_{n}(x)$ can be interpreted as recoded versions of $FSD(x,P)$ and $FSD_{n}(x)$, respectively.

This interpretation of a local measure as a recoded version of a global measure applies also to the kernel-based HMD (\citeauthor{CueFebFra2006} \citeyear{CueFebFra2006}). The sample HMD is given by

\begin{equation*}
\label{eq:hmd}
HMD_{n}(x) = \sum_{y \in (y_{i})_{i=1, \ldots, n}} \kappa(x,y).
\end{equation*}

\noindent Now, let the sample inner products sum function at $x$ be

\begin{equation*}
\label{eq:ips}
IPS_{n}(x) = \sum_{y \in (y_{i})_{i=1, \ldots, n}} \langle x,y\rangle.
\end{equation*}

\noindent Then, $HMD_{n}(x)$ can be interpreted as a recoded version of $IPS_{n}(x)$, that is,

\begin{equation}
\label{eq:hmdToIps}
HMD_{n}(x) = \sum_{\phi(y) \in (\phi(y_{i}))_{i=1, \ldots, n}} \langle \phi(x), \phi(y) \rangle = IPS_{n}(\phi(x)),
\end{equation}

\noindent where $IPS_{n}(\phi(x))$ is the inner products sum function at
$\phi(x)$ relative to the recoded sample
$(\phi(y_{i}))_{i=1,\ldots, n}$. Therefore, (\ref{eq:S02_19}), (\ref{eq:funKerDep}) and (\ref{eq:hmdToIps}) show that the idea of local depth can be expressed by means of the mathematical notion of embedding map.

\section{DEPTH-BASED SUPERVISED CLASSIFICATION FOR FUNCTIONAL DATA}
\label{sec:classification}

In supervised functional classification, the natural theoretical framework is given by the random pair $(Y,G)$, where $Y$ is a functional random variable and $G$ is a categorical random variable describing the class membership. From now on, we assume that $G$ takes values $0$ or $1$ and that we observe a sample of $n$ independent pairs taken from the distribution of $(Y,G)$, i.e., $(y_{i}, g_{i})_{i=1,\ldots, n}$, where $n_0$ observations come from the group with label 0, $n_1$ observations come from the group with label 1 and $n=n_0+n_1$. Additionally, we observe an independent curve $x$ identically distributed as $Y$, but with unknown class membership. Using the information contained in $(y_{i},g_{i})_{i=1, \ldots, n}$, the goal of any supervised functional classification method is to provide a rule to classify the curve $x$ (\citeauthor*{FerVie2006} \citeyear{FerVie2006}).

Several supervised functional classification methods have been proposed in the literature. For instance, \citet*{HasBujTib1995} have proposed a penalized version of the multivariate linear discriminant analysis technique, whereas \citet*{JamHas2001} have directly built a functional linear discriminant analysis procedure that uses natural cubic spline functions to model the observations. Using a P-spline approach, \citet*{MarEil1999} have considered functional supervised classification as a special case of a generalized linear regression model. \citet*{HalPosPre2001} have suggested to perform dimension reduction by means of functional principal component analysis and then to solve the derived multivariate problem with quadratic discriminant analysis or kernel methods. \citet*{FerVie2003} have developed a functional kernel-type classifier. \citet*{Epi2008} have proposed to describe curves by means of shape feature vectors and to use classical multivariate classifiers for the discrimination stage. Finally, \citet*{BiaBunWeg2005} and \cite{CerGuy2006} have studied some consistency properties of the extension of the \textit{k}-nearest neighbor procedure to infinite-dimensional spaces. The first extension considers a reduction of the dimensionality of the regressors based on a Fourier basis system, and the second generalization deals with the real infinite dimension of the spaces under consideration. Note that the \textit{k}-nearest neighbor method is indeed a general tool that can be used to perform functional nonparametric regression, and that classification is a specific case that occurs when the response of the regression model is categorical (for more details and theoretical results on the general case, see \citeauthor{BurFerVie2009} \citeyear{BurFerVie2009} and \citeauthor{KudVie2013} \citeyear{KudVie2013}).

The main difference between the above-mentioned methods and the depth-based procedures, DTM, WAD and WMD, is that the second ones are specially designed
for datasets that may contain outlying curves. For this reason, in Section \ref{sec:simStudy} we carry out a simulation study with scenarios that allow outliers, whereas in Section \ref{sec:real} we consider two potentially contaminated real datasets. Let us briefly describe
DTM, WAD and WMD: first, the distance to the trimmed mean procedure
(DTM) computes the
$\alpha$-trimmed mean $m_{g}^{\alpha}$, i.e., the mean of the
$1-\alpha$ deepest curves of each group, where
$\alpha$ is a certain proportion, and it assigns $x$ to the group for which
$\|x-m_{g}^{\alpha}\|$ is less. Clearly, the contribution of the
chosen functional depth is at the trimming stage, and it allows to
obtain robust means (\citeauthor*{LopRom2006} \citeyear{LopRom2006}). Second, the weighted averaged distance procedure (WAD) computes, for
each group, a weighted average of the distances $\|x-y_{i}\|_{i=1,
\ldots, n_{g}}$, where the weights are given by the within-group depth
values, $D(y_{i}; g=\cdot)_{i=1, \ldots, n_{g}}$, and it assigns $x$ to the
group for which the weighted averaged distance is less (\citeauthor*{LopRom2006} \citeyear{LopRom2006}). Finally,
the within maximum depth procedure (WMD) computes the depth value of the curve
$x$ with respect to each group, and
it assigns $x$ to the group for which the depth value is higher (\citeauthor{CueFebFra2007}
\citeyear{CueFebFra2007}).

Any functional depth can be used to perform supervised functional
classification together with any of the depth-based methods described above. In
Sections \ref{sec:simStudy} and \ref{sec:real} we compare the
performances of DTM, WAD and WMD when used in conjunction with
alternative functional depths such as FMD, HMD, RTD, IDD, MBD, FSD and KFSD introduced in Sections \ref{sec:intro} and \ref{sec:FSDs}. The \textit{k}-nearest neighbor procedure (\textit{k}-NN) is used as
benchmark in our simulation and real data studies. Its
generalization to infinite-dimensional spaces consists in the following
rule: look at the \textit{k} nearest neighbors of $x$ among
$(y_{i})_{i=1, \ldots, n}$, and choose its group according to the
majority vote. The search of the neighbors is based on the norm
defined on $\mathbb{H}$. It is worth noting that the
classification rule characterizing \textit{k}-NN makes the method
rather robust to the presence of outliers, which is the reason why  \textit{k}-NN represents an interesting competitor for DTM, WAD and WMD.

\section{SIMULATION STUDY}
\label{sec:simStudy}

In Sections \ref{sec:FSDs} and \ref{sec:classification} we have
presented three different depth-based classification procedures (DTM, WAD and
WMD) and seven different functional  depths (FMD, HMD, RTD, IDD,
MBD, FSD and KFSD). Pairing all the procedures with all the
depths, we obtain $21$ depth-based classification methods, plus \textit{k}-NN. The goal of this section is to compare them through an extensive simulation study. From now on, we refer to each method
by the notation procedure+depth: for example,
DTM+FMD refers to the method obtained by using DTM together with
FMD. 

We mainly explore the effectiveness of the methods in supervised
classification problems in which the appropriate class membership
of the curves is hard to be deduced by using graphical tools
and/or in scenarios in which outlying curves are allowed. In these cases it may happen that the curve we want to classify is geometrically rather central relative to the two training samples, but also relatively far from one of them, although not in an obvious way. In such scenarios, the use of a local depth may be a better strategy than the use of a global depth when it is considered a depth-based classification method.

Both classification methods and depths may depend on some
parameters or assumptions. Regarding the methods, DTM depends on the
trimming parameter $\alpha$, that we set at $\alpha=0.2$, as in \cite{LopRom2006}. For the benchmark procedure \textit{k}-NN, we take $k=5$ nearest neighbors since it is a standard choice and the method is reasonably robust with respect to the chosen
$k$. Regarding the functional depths, for HMD, following the
recommendations made by \citet{FebGalGon2008}, we choose the $L^{2}$ norm as norm function, and the positive Gaussian kernel $\kappa(x,y) =
(2/\sqrt{2\pi})\times\exp (-\|x-y\|^2/2\sigma^2)$, with bandwidth $\sigma$ equal to the 15th percentile of
the empirical distribution of $\{\|y_{i}-y_{j}\|, i,j=1, \ldots,
n\}$, as kernel
function. Note that for WMD+HMD we need to use a normalized version of
HMD to make its range equal to $[0,1]$. For RTD and IDD, we
consider $p=50$ random projections and generate the random
directions through a Gaussian process. For MBD, we set the maximum
number of curves defining each band to $J=2$. For FSD and KFSD, we use the $L^{2}$ norm as norm function. In addition, for KFSD we use a Gaussian kernel function, and in particular a functional version of the one
used by \citet{CheDanPenBar2009}, i.e.,

\begin{equation*}
\label{eq:kappaKFSD}
\kappa(x,y) = \exp \left(-\frac{\|x-y\|^2}{\sigma^2}\right).
\end{equation*}

\noindent With the aim of exploring
the effects of the choice of the bandwidth $\sigma$ on the performances of
KFSD, we consider a set of representative percentiles of the empirical distribution of $\{\|y_{i}-y_{j}\|, i,j=1, \ldots, n\}$, i.e, the 15th, 25th, 33rd, 50th, 66th, 75th and 85th percentile, thus obtaining 7 different $\sigma$, and 7 different versions of KFSD. Note that when KFSD is used with low percentiles, it considers small neighborhoods and can be viewed as a potential functional density estimator. On the other hand, when KFSD is used with large percentiles, it considers wide neighborhoods and behaves as a global-oriented functional depth.

As aforementioned, we are mainly interested in challenging classification scenarios. Indeed, we overlook problems in which the curves may be almost well classified by a preliminary graphical analysis and focus on classification scenarios in which the differences among groups are hard to be detected graphically. Moreover, we allow outliers in one part of the simulation study. We consider two-groups scenarios throughout the whole simulation study: $g \in \{0, 1\}$ is the label assigned to each group and $x_{g}(t)$ is the curve generating process for group $g$.

In absence of contamination, we initially consider two different pairs of curve generating processes:

\begin{compactenum}

\item First pair of curve generating processes (from now on, CGP1) with $t \in [0,1]$

\begin{equation*}\label{eq:CGP1}
\begin{array}{ll}
x_{0}(t) &= 4t + \epsilon(t), \\
x_{1}(t) &= 8t-2 + \epsilon(t),
\end{array}
\end{equation*}

\noindent where $\epsilon(t)$ is a zero-mean Gaussian component with covariance function given by

\begin{equation}
\label{eq:cov01}
\mathbb{E}(\epsilon(t),\epsilon(s)) = 0.25 \exp{(-(t-s)^2)}, \quad t,s \in [0, 1].
\end{equation}

\item Second pair of curve generating processes (from now on, CGP2) with $t \in [0,2\pi]$

\begin{equation}\label{eq:CGP2}
\begin{array}{ll}
x_{0}(t) &= u_{01}\sin t+u_{02}\cos t, \\
x_{1}(t) &= u_{11}\sin t+u_{12}\cos t,
\end{array}
\end{equation}

\noindent where $u_{01}$ and $u_{02}$ are i.i.d. observations from a continuous uniform random variable between 0.05 and 0.1, whereas $u_{11}$ and $u_{12}$ are i.i.d. observations from a continuous uniform random variable between 0.1 and 0.12.

\end{compactenum}

The main difference between CGP1 and CGP2 is that for
CGP1 both $x_{0}(t)$ and $x_{1}(t)$ are composed of deterministic
and linear mean functions, plus a random component, while for CGP2 both
$x_{0}(t)$ and $x_{1}(t)$ are exclusively composed of random and
nonlinear mean functions.

\begin{figure}[!htb]
\centering
\includegraphics[scale=0.5]{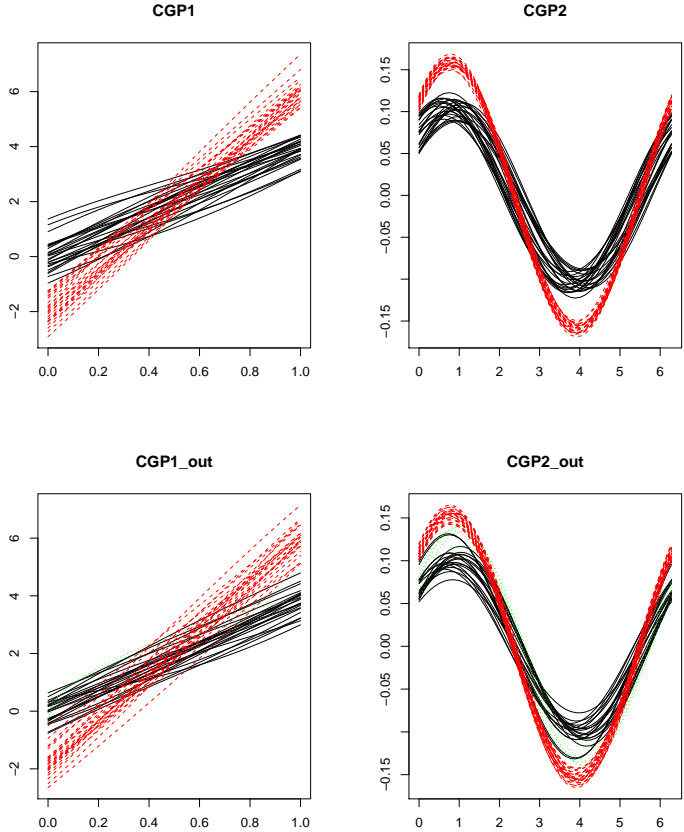}
\captionsetup{width=0.8\textwidth}
\caption{Simulated datasets from CGP1, CGP2, CGP1$_{out}$ and CGP2$_{out}$: each dataset contains 25 curves from group $g=0$ and 25 dashed curves from group $g=1$. For CGP1 and CGP2, the curves from group $g=0$ are all noncontaminated (solid). For CGP1$_{out}$ and CGP2$_{out}$, the curves from group $g=0$ can be noncontaminated (solid) or contaminated (dotted).}
\label{fig:PM1PM2C1C2}
\end{figure}

To allow contamination, we consider the following modified
version of CGP1 and CGP2, where the contamination affects only
group 0:
\begin{compactenum}
\item First pair of curve generating processes allowing outliers (from now on, CGP1$_{out}$) with $t \in [0,1]$

\begin{equation*}\label{eq:CGP1out}
\begin{array}{ll}
x_{0}(t) &=  \left\{
\begin{array}{ll}
4t + \epsilon(t), &\mbox{ with probability } 1-q,\\
4\sqrt{t} + \epsilon(t), &\mbox{ with probability } q.
\end{array} \right.
\\
x_{1}(t) &= 8t-2 + \epsilon(t),
\end{array}
\end{equation*}

\noindent where $\epsilon(t)$ is a zero-mean Gaussian component with covariance function given by (\ref{eq:cov01}) and $0<q<1$.

\item Second pair of curve generating processes allowing outliers (from now on, CGP2$_{out}$) with $t \in [0,2\pi]$

\begin{equation*}\label{eq:CGP2_out}
\begin{array}{ll}
x_{0}(t) &=  \left\{
\begin{array}{ll}
u_{01}\sin t+u_{02}\cos t, &\mbox{ with probability } 1-q,\\
u_{01}\sin t+u_{12}\cos t, &\mbox{ with probability } q.
\end{array} \right.
\\
x_{1}(t) &= u_{11}\sin t+u_{12}\cos t,
\end{array}
\end{equation*}

\noindent where $u_{i,j},\, i=0,1, \, j=1,2$ and $u_{02}$ are defined as for CGP2 and $0<q<1$.
\end{compactenum}
In Figure \ref{fig:PM1PM2C1C2} we report a simulated dataset from CGP1, CGP2, CGP1$_{out}$ and CGP2$_{out}$.

Next, we present the details of the simulation study: for each model, we generate 125 replications. For CGP1$_{out}$ and CGP2$_{out}$, we set the contamination probability $q= 0.10$. For each replication, we generate 100 curves, 50 for $g=0$ and 50 for $g=1$. We use 25 curves from $g=0$ and 25 curves from $g=1$ to build each training sample, and we classify the remaining curves. All curves are generated using a discretized and finite set of 51 equidistant points between 0 and 1 or 0 and 2$\pi$, depending on the model. For all the functional depths, we use a discretized version of their definitions. We perform the comparison among methods in terms of misclassification percentages and we report their means and standard deviations in the next tables. 

For what concerns KFSD, since we consider 7 different percentiles to set $\sigma$, for each replication and a given classification method, we base the choice of the percentile on a cross validation step that we organize as follows:

\begin{compactenum}
\item We divide each initial training sample in 5 groups-balanced cross validation test samples, each one of size 10, and we pair each of them with its natural cross validation training sample composed of the remaining 40 curves.

\item Then, for a given replication, we have available five pairs of cross validation training and test samples. For each pair and a given classification method, we classify each curve in the test sample using the 7 different percentiles.

\item For a given replication and classification method, once we obtain the 7 misclassification errors, we search for the minimum misclassification error and its corresponding percentile.

\item Finally, we use the percentile selected in 3. to perform the classification for the pair of initial training and test samples. 
\end{compactenum}
 
In a preliminary stage of the study, we have observed that it is common to have ties in the cross validation step described above. For this reason, in presence of ties, we propose to break them using second criteria based on the additional information provided by the different classification methods: for DTM, we select the percentile that minimizes the sum of the distances of the curves in the cross validation test samples to the $\alpha$-trimmed mean of the group at which the curves belong; for WAD, we select the percentile that minimizes the sum of the weighted averaged distances of the curves in the cross validation test samples to the group at which the curves belong; for WMD, we select the percentile that maximizes the sum of the scaled KFSD of the curves in the cross validation test samples when included in the group at which they belong. The second cross validation criteria break almost all the ties; however, if after considering the second criterion a tie is not broken, we propose to break it randomly.

To give an idea about the computational issue, note that for both DTM and WAD the key step consists in computing the within-group depth values of the training curves. Hence, we report the computational times (in seconds) that require the considered functional depths to perform this task for a training sample of size 50: 0.02 for FMD, 0.08 for HMD, 0.02 for RTD, 0.02 for IDD, less than 0.01 for MBD, 0.05 for FSD, and 0.14 for KFSD\footnote{For FMD, HMD, RTD and IDD we have used the corresponding \texttt{R} functions that are available in the \texttt{R} package \texttt{fda.usc} on \texttt{CRAN} (\citeauthor{FebOvi2012} \citeyear{FebOvi2012}); for MBD we have followed the guidelines contained in \citet*{SunGenNyc2012}; for FSD and KFSD we have built some functions for \texttt{R}, which are available upon request. Features of the workstation: Intel Core i7-3.40GHz and 16GB of RAM.}. Additionally, a 7-percentiles KFSD analysis takes 0.54 seconds, and a 7-percentiles KFSD analysis combined with a 5-subsamples cross validation step takes 1.70 seconds. Therefore, the computation of KFSD is widely feasible, and also the option of searching for an optimal percentile by means of a cross validation step does not cause major computational problems, whereas it will bring some classification benefits.

Tables \ref{tab:PM1}-\ref{tab:C2} report the performances that we have observed with the functional data generated from CGP1, CGP2, CGP1$_{out}$ and CGP2$_{out}$. 

Regarding the KFSD-based methods, for given initial training and test samples and classification method, it may happen to observe the same misclassification error at the 7 different percentiles. In such cases, the cross validation step turns out to be unnecessary. On the contrary, the cross validation is required if there are at least two different misclassification errors. For each model and method, Table \ref{tab:percentilesCGPs_A} reports the percentages of replications for which the KFSD cross validation step is required. In the same table we also report the best performing percentiles for the pairs of initial training and test samples since this information permits to identify the type of local analysis that would classify best.

\begin{table}[!htbp]
\parbox{.475\textwidth}{
\captionsetup{justification=justified,width=.475\textwidth}
\caption{CGP1. Means and standard deviations \textit{(in parenthesis)} of the misclassification percentages for DTM, WAD and WMD-based methods and \textit{k}-NN.}
\centering
\scalebox{0.55}{
\begin{tabular}{cccccccc}
\hline
\hline
Method/Depth & FMD & HMD & RTD & IDD & MBD & FSD & KFSD\\
\hline
\multirow{2}{*}{DTM} & 0.16 & 0.22 & 0.14 & 0.14 & 0.18 & 0.19 & 0.14\\
& \textit{(0.65)} & \textit{(0.73)} & \textit{(0.58)} & \textit{(0.63)} & \textit{(0.62)} & \textit{(0.64)} & \textit{(0.58)} \\
 \hline
\multirow{2}{*}{WAD} & 0.10 & 0.16 & 0.11 & 0.11 & 0.08 & 0.13 & 0.10\\
& \textit{(0.50)} & \textit{(0.60)} & \textit{(0.53)} & \textit{(0.53)} & \textit{(0.47)} & \textit{(0.55)} & \textit{(0.50)}\\
\hline
\multirow{2}{*}{WMD} & 15.09 & 1.66 & 21.90 & 18.06 & 11.82 & 3.30 & 0.13\\
& \textit{(5.43)} & \textit{(2.46)} & \textit{(6.82)} & \textit{(6.27)} & \textit{(4.95)} & \textit{(2.89)} & \textit{(0.66)}\\
\hline
\hline
\multirow{2}{*}{\textit{k}-NN} & \multicolumn{7}{c}{0.11} \\
& \multicolumn{7}{c}{\textit{(0.46)}}\\
\hline
\end{tabular}}
\label{tab:PM1}
}
\hfill
\parbox{.475\textwidth}{
\captionsetup{justification=justified,width=0.475\textwidth}\caption{CGP2. Means and standard deviations \textit{(in parenthesis)} of the misclassification percentages for DTM, WAD and WMD-based methods and \textit{k}-NN.}
\centering
\scalebox{0.55}{
\begin{tabular}{cccccccc}
\hline
\hline
Method/Depth & FMD & HMD & RTD & IDD & MBD & FSD & KFSD\\
\hline
\multirow{2}{*}{DTM} & 2.43 & 2.59 & 2.40 & 2.45 & 2.45 & 2.42 & 2.35\\
& \textit{(2.01)} & \textit{(2.24)} & \textit{(2.00)} & \textit{(2.05)} & \textit{(1.98)} & \textit{(1.99)} & \textit{(2.10)}\\
\hline
\multirow{2}{*}{WAD} & 3.38 & 3.33 & 3.10 & 3.22 & 3.20 & 3.02 & 3.14\\
& \textit{(2.32)} & \textit{(2.35)} & \textit{(2.25)} & \textit{(2.33)} & \textit{(2.23)} & \textit{(2.19)} & \textit{(2.25)} \\
\hline
\multirow{2}{*}{WMD} & 0.10 & 2.16 & 1.12 & 0.99 & 0.13 & 0.14 & 0.11\\
& \textit{(0.43)} & \textit{(2.75)} & \textit{(1.99)} & \textit{(1.75)} & \textit{(0.49)} & \textit{(0.52)} & \textit{(0.53)} \\
\hline
\hline
\multirow{2}{*}{\textit{k}-NN} & \multicolumn{7}{c}{0.88} \\
& \multicolumn{7}{c}{\textit{(1.53)}}\\
\hline
\end{tabular}}
\label{tab:PM2}
}
\end{table}

\begin{table}[!htbp]
\parbox{.475\textwidth}{
\captionsetup{justification=justified,width=.475\textwidth}
\caption{CGP1$_{out}$. Means and standard deviations \textit{(in parenthesis)} of the misclassification percentages for DTM, WAD and WMD-based methods and \textit{k}-NN.}
\centering
\scalebox{0.55}{
\begin{tabular}{cccccccc}
\hline
\hline
Method/Depth & FMD & HMD & RTD & IDD & MBD & FSD & KFSD\\
\hline
\multirow{2}{*}{DTM} & 0.11 & 0.08 & 0.08 & 0.08 & 0.06 & 0.11 & 0.05\\
& \textit{(0.46)} & \textit{(0.39)} & \textit{(0.39)} & \textit{(0.39)} & \textit{(0.35)} & \textit{(0.46)} & \textit{(0.31)} \\
\hline
\multirow{2}{*}{WAD} & 0.06 & 0.06 & 0.06 & 0.08 & 0.06 & 0.06 & 0.06\\
& \textit{(0.35)} & \textit{(0.35)} & \textit{(0.35)} & \textit{(0.39)} & \textit{(0.35)} & \textit{(0.35)} & \textit{(0.35)} \\
\hline
\multirow{2}{*}{WMD} & 14.46 & 2.34 & 22.93 & 18.93 & 11.47 & 3.26 & 0.08\\
& \textit{(5.54)} & \textit{(2.97)} & \textit{(6.92)} & \textit{(7.11)} & \textit{(5.02)} & \textit{(2.80)} & \textit{(0.39)} \\
\hline
\hline
\multirow{2}{*}{\textit{k}-NN} & \multicolumn{7}{c}{0.08} \\
& \multicolumn{7}{c}{\textit{(0.39)}}\\
\hline
\end{tabular}}
\label{tab:C1}
}
\hfill
\parbox{.475\textwidth}{
\captionsetup{justification=justified,width=.475\textwidth}
\caption{CGP2$_{out}$. Means and standard deviations \textit{(in parenthesis)} of the misclassification percentages for DTM, WAD and WMD-based methods and \textit{k}-NN.}
\centering
\scalebox{0.55}{
\begin{tabular}{cccccccc}
\hline
\hline
Method/Depth & FMD & HMD & RTD & IDD & MBD & FSD & KFSD\\
\hline
\multirow{2}{*}{DTM} & 3.87 & 4.11 & 3.84 & 3.81 & 3.76 & 3.74 & 3.71\\
& \textit{(2.86)} & \textit{(3.11)} & \textit{(2.84)} & \textit{(2.75)} & \textit{(2.86)} & \textit{(2.83)} & \textit{(2.91)} \\
\hline
\multirow{2}{*}{WAD} & 4.94 & 4.94 & 4.67 & 4.74 & 4.80 & 4.48 & 4.70\\
& \textit{(3.36)} & \textit{(3.44)} & \textit{(3.25)} & \textit{(3.25)} & \textit{(3.24)} & \textit{(3.16)} & \textit{(3.29)} \\
\hline
\multirow{2}{*}{WMD} & 0.82 & 2.90 & 4.19 & 3.84 & 0.83 & 0.83 & 0.58\\
& \textit{(1.48)} & \textit{(3.31)} & \textit{(3.72)} & \textit{(3.26)} & \textit{(1.55)} & \textit{(1.46)} & \textit{(1.29)} \\
\hline
\hline
\multirow{2}{*}{\textit{k}-NN} & \multicolumn{7}{c}{1.97} \\
& \multicolumn{7}{c}{\textit{(2.05)}}\\
\hline
\end{tabular}}
\label{tab:C2}
}
\end{table}

\begin{table}[!htbp]
\captionsetup{justification=justified,width=0.75\textwidth}
\caption{DTM+KFSD, WAD+KFSD and WMD+KFSD, and curve generating processes CGP1, CGP2, CGP1$_{out}$ and CGP2$_{out}$: percentages of the replications for which cross validation is required and best performing percentiles for the initial training and test samples.}
\centering
\scalebox{0.60}{
\begin{tabular}{c|c|c|c|c|c|c|c}
\hline
\hline
Model & Method & Percentage & Percentiles & Model & Method & Percentage & Percentiles\\
\hline
\multirow{3}{*}{CGP1} & DTM & 7.20 & 50th & \multirow{3}{*}{CGP2} & DTM & 21.60 & 66th\\ 
& WAD & 2.40 & 15th, 25th, 50th, 66th, 75th & & WAD & 13.60 & 85th\\ 
& WMD & 56.00 & 50th, 66th, 75th & & WMD & 12.80 & 66th, 85th\\ 
\hline
\multirow{3}{*}{CGP1out} & DTM & 3.20 & 15th, 25th, 50th, 66th & \multirow{3}{*}{CGP2out} & DTM & 26.40 & 85th\\ 
& WAD & 0.00 & all & & WAD & 19.20 & 85th\\ 
& WMD & 58.40 & 66th & & WMD & 37.60 & 75th\\ 
\hline 
\hline
\end{tabular}
\label{tab:percentilesCGPs_A}
}
\end{table}

The results in Tables \ref{tab:PM1}-\ref{tab:percentilesCGPs_A} show that:

\begin{compactenum}

\item When the curves are generated from CGP1 or CGP1$_{out}$, WAD
is the best classification procedure, but also the
performances of DTM are acceptable. Both procedures turn out rather stable with respect to the choice of the
functional depth. On the contrary, WMD is more sensitive to the choice of the depth measure, and only the combination WMD+KFSD is able to compete with WAD and DTM. The performances of \textit{k}-NN are quite good, but they
are not the best ones neither for CGP1 nor for
CGP1$_{out}$. Indeed, for CGP1, the best method is WAD+MBD
(0.08\%), whereas WAD+KFSD and WAD+FMD are the second
best methods (0.10\%). For CGP1$_{out}$, the best method is DTM+KFSD (0.05\%), and some other spatial depth-based methods perform also quite good, e.g., WAD+FSD and WAD+KFSD (0.06\%). 
Finally, note that WMD+KFSD behaves reasonably well in both scenarios (0.13\% with CGP1 and 0.08\% with CGP1$_{out}$).

\item When the curves are generated from CGP2 or CGP2$_{out}$, WMD, in conjunction with FMD, MBD, FSD and KFSD, is clearly the best performing classification procedure: indeed, the four resultant methods markedly outperform \textit{k}-NN. For CGP2, the best method is WMD+FMD (0.10\%), but the performance of WMD+KFSD is almost equal (0.11\%). For CGP2$_{out}$, the best method is clearly WMD+KFSD (0.58\%).

\item The cross validation step is in general required more by WMD+KFSD, and to a smaller extent by DTM+KFSD, and finally by WAD+KFSD. For example, with CGP1$_{out}$ the cross validation step is completely unnecessary for WAD+KFSD, whereas it is advisable for more than one-half of the replications with WMD+KFSD. On the other hand, looking at the best performing percentiles and focusing on the methods highlighted in the previous two points, we observe the following: under CGP1 and CGP1$_{out}$, WAD+KFSD reaches its best performances with several percentiles; under CGP2 and CGP2$_{out}$, the best performances of WMD+KFSD are with rather high percentiles. This last result is coherent with the good performances of WMD+FSD under CGP2 and CGP2$_{out}$. Indeed, the higher is the percentile, the less local-oriented is KFSD, and its behavior tends towards the behavior of FSD, its global-oriented counterpart.

\end{compactenum}

Observing the curves generated from CGP1 in Figure \ref{fig:PM1PM2C1C2}, we can appreciate a rather strong data dependence structure, which is due to the covariance function in (\ref{eq:cov01}).
On the other hand, observing the curves generated from CGP2 at any fixed $t \in [0, 2\pi]$, we can appreciate low variability in the data. We enhance our simulation study by relaxing these two features of CGP1 and CGP2 and considering two modifications of them: first, we consider a variation of CGP1 (from now on, CGP3) which consists in substituting the covariance function of the additive zero-mean Gaussian component previously given by (\ref{eq:cov01}) with a weaker version defined as

\begin{equation*}
\label{eq:cov01Mod}
\mathbb{E}(\epsilon(t),\epsilon(s)) = 0.30 \exp{(-|t-s|/0.3)}, \quad t,s \in [0, 1].
\end{equation*}

\noindent Second, we consider a variation of CGP2 (from now on, CGP4) which consists in adding to the two processes in (\ref{eq:CGP2}) two identical additive zero-mean Gaussian components having covariance function

\begin{equation*}
\label{eq:cov02Mod}
\mathbb{E}(\epsilon(t),\epsilon(s)) = 0.00025 \exp{(-(t-s)^2)}, \quad t,s \in [0, 2\pi].
\end{equation*}

\begin{figure}[!htb]
\centering
\includegraphics[scale=0.5]{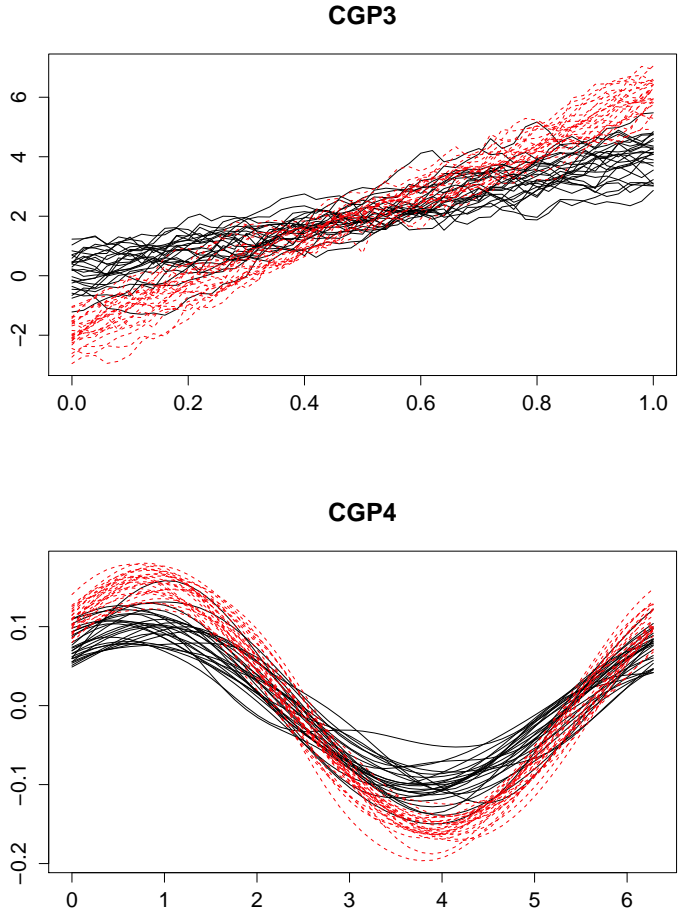}
\captionsetup{width=0.8\textwidth}
\caption{Simulated datasets from CGP3 and CGP4: each dataset contains 25 solid curves from group $g=0$ and 25 dashed curves from group $g=1$.}
\label{fig:CGP3_CGP4}
\end{figure}

Figure \ref{fig:CGP3_CGP4} reports two simulated datasets from CGP3 and CGP4. We use these models to develop the third and last part of the simulation study. Tables \ref{tab:CGP3} and \ref{tab:CGP4} report the performances of the 21 depth-based methods and of \textit{k}-NN, whereas Table \ref{tab:percentilesCGPs_B} is the analogous of Table \ref{tab:percentilesCGPs_A} for CGP3 and CGP4.

\begin{table}[!htb]
\parbox{.475\textwidth}{
\captionsetup{justification=justified,width=.475\textwidth}
\caption{CGP3. Means and standard deviations \textit{(in parenthesis)} of the misclassification percentages for DTM, WAD and WMD-based methods and \textit{k}-NN.}
\centering
\scalebox{0.55}{
\begin{tabular}{cccccccc}
\hline
\hline
Method/Depth & FMD & HMD & RTD & IDD & MBD & FSD & KFSD\\
\hline
\multirow{2}{*}{DTM} & 1.33 & 1.36 & 1.39 & 1.36 & 1.36 & 1.38 & 1.31\\
& \textit{(1.39)} & \textit{(1.40)} & \textit{(1.44)} & \textit{(1.47)} & \textit{(1.31)} & \textit{(1.42)} & \textit{(1.35)}\\
\hline
\multirow{2}{*}{WAD} & 1.14 & 1.18 & 1.20 & 1.17 & 1.17 & 1.20 & 1.17\\
& \textit{(1.33)} & \textit{(1.32)} & \textit{(1.37)} & \textit{(1.35)} & \textit{(1.32)} & \textit{(1.34)} & \textit{(1.32)}\\
\hline
\multirow{2}{*}{WMD} & 5.26 & 2.93 & 17.42 & 14.64 & 4.29 & 1.44 & 1.20\\
& \textit{(3.12)} & \textit{(2.66)} & \textit{(5.90)} & \textit{(5.51)} & \textit{(2.76)} & \textit{(1.56)} & \textit{(1.39)}\\
\hline
\hline
\multirow{2}{*}{\textit{k}-NN} & \multicolumn{7}{c}{1.39} \\
& \multicolumn{7}{c}{\textit{(1.46)}}\\
\hline
\end{tabular}}
\label{tab:CGP3}
}
\hfill
\parbox{.475\textwidth}{
\captionsetup{justification=justified,width=.475\textwidth}
\caption{CGP4. Means and standard deviations \textit{(in parenthesis)} of the misclassification percentages for DTM, WAD and WMD-based methods and \textit{k}-NN.}
\centering
\scalebox{0.55}{
\begin{tabular}{cccccccc}
\hline
\hline
Method/Depth & FMD & HMD & RTD & IDD & MBD & FSD & KFSD\\
\hline
\multirow{2}{*}{DTM} & 2.38 & 2.53 & 2.30 & 2.38 & 2.26 & 2.30 & 2.26\\
& \textit{(2.61)} & \textit{(2.59)} & \textit{(2.46)} & \textit{(2.47)} & \textit{(2.42)} & \textit{(2.45)} & \textit{(2.40)} \\
\hline
\multirow{2}{*}{WAD} & 3.55 & 3.46 & 3.46 & 3.46 & 3.60 & 3.36 & 3.41\\
& \textit{(3.08)} & \textit{(3.08)} & \textit{(2.87)} & \textit{(2.98)} & \textit{(3.07)} & \textit{(2.92)} & \textit{(2.94)}\\
\hline
\multirow{2}{*}{WMD} & 15.52 & 2.88 & 28.54 & 24.02 & 12.96 & 4.90 & 0.82\\
& \textit{(5.29)} & \textit{(3.19)} & \textit{(6.99)} & \textit{(6.98)} & \textit{(4.76)} & \textit{(3.96)} & \textit{(1.57)}\\
\hline
\hline
\multirow{2}{*}{\textit{k}-NN} & \multicolumn{7}{c}{1.81} \\
& \multicolumn{7}{c}{\textit{(2.28)}}\\
\hline
\end{tabular}}
\label{tab:CGP4}
}
\end{table}

\begin{table}[!htb]
\captionsetup{justification=justified,width=0.75\textwidth}
\caption{DTM+KFSD, WAD+KFSD and WMD+KFSD, and curve generating processes CGP3 and CGP4: percentages of the replications for which cross validation is required and best performing percentiles for the initial training and test samples.}
\centering
\scalebox{0.60}{
\begin{tabular}{c|c|c|c|c|c|c|c}
\hline
\hline
Model & Method & Percentage & Percentiles & Model & Method & Percentage & Percentiles\\
\hline
\multirow{3}{*}{CGP3} & DTM & 12.00 & 25th, 75th, 85th & \multirow{3}{*}{CGP4} & DTM & 14.40 & 85th\\ 
& WAD & 0.80 & 25th, 33rd, 50th, 66th, 75th & & WAD & 7.20 & 66th\\ 
& WMD & 28.00 & 75th, 85th & & WMD & 77.60 & 33rd\\ 
\hline 
\hline
\end{tabular}
\label{tab:percentilesCGPs_B}
}
\end{table}

The results in Tables \ref{tab:CGP3}-\ref{tab:percentilesCGPs_B} show that:

\begin{compactenum}

\item When the curves are generated from CGP3, which is a modification of CGP1, we indeed observe similar results. WAD is the best classification procedure, but the behavior of DTM is not bad at all. WMD heavily fails, with the exceptions of the spatial depths KFSD and FSD, and HMD. \textit{k}-NN is a competitive classification procedure, but all the DTM and WAD-based methods, and WMD+KFSD outperform it. The best method is WAD+FMD (1.14\%), whereas there are several best second methods (1.17\%), including the spatial method WAD+KFSD.

\item When the curves are generated from CGP4, which is a modification of CGP2, we observe that WMD+KFSD is the only method able to outperform \textit{k}-NN (0.82\% against 1.81\%), whereas the remaining methods highlighted for CGP2, i.e., WMD+FMD, WMD+MBD and WMD+FSD, drastically worsen.

\item As for the previous models, the cross validation step is in general required more by WMD+KFSD, and to a smaller extent by DTM+KFSD, and finally by WAD+KFSD. For example, with CGP3 the cross validation step is almost unnecessary for WAD+KFSD, whereas it is advisable for more than one-quarter of the replications with WMD+KFSD. On the other hand, looking at the best performing percentiles and focusing on the methods highlighted in the previous two points, we observe that for CGP3 WAD+KFSD reaches its best performance with all the percentiles except the 15th one, whereas for CGP4 the best performances of WMD+KFSD are with the 33rd percentile, which means that for this method there is a gain when a rather strong local approach is implemented.

\end{compactenum}

To conclude, we have observed that for the curve generating processes having a deterministic and linear mean function and a random component, i.e., CGP1, CGP1$_{out}$ and CGP3, WAD+KFSD is among the best and most stable classification methods, and both DTM+KFSD and WMD+ KFSD have performances that are not so different. On the other hand, for the curve generating processes having a random and nonlinear mean function, i.e., CGP2, CGP2$_{out}$ and CGP4, WMD+KFSD is clearly the best classification method. Therefore, KFSD-based functional supervised classification is certainly a good option to discriminate curves.

\section{REAL DATA STUDY}
\label{sec:real}
To complete the comparison among the depth-based methods and \textit{k}-NN, we also consider two real datasets.

\subsection{GROWTH DATA}
\label{subsec:growth}
\begin{figure}[!htb]
\centering
\includegraphics[scale=0.5]{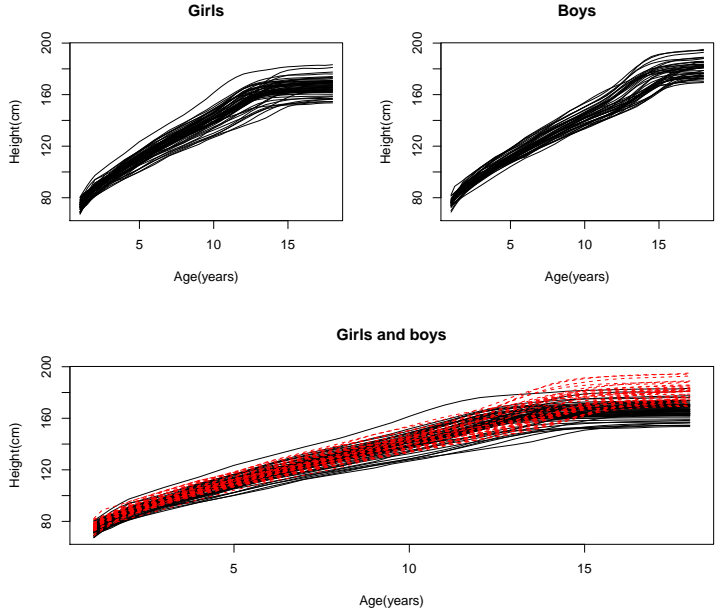}
\captionsetup{width=0.8\textwidth}
\caption{Growth curves: 54 heights of girls (top left), 39 heights of boys (top right), 93 heights of girls and boys (bottom; the dashed curves are boys).}
\label{fig:growth}
\end{figure}

The first real dataset consists of $93$ growth curves: $54$ are heights of girls, $39$ are heights of boys. All of them are observed at a common discretized set of 31 nonequidistant ages between 1 and 18 years. Figure \ref{fig:growth} shows the curves (for more details about this dataset, see \citeauthor*{RamSil2005} \citeyear{RamSil2005}).

We use natural cubic spline interpolation to estimate the growth curves at a common and equally spaced domain. Clearly, other techniques can be used for this task, but we choose this standard interpolation technique before we focus on classification.

This dataset has been analyzed by \cite{LopRom2006} and \citet{CueFebFra2007}. From our point of view, these data are interesting mainly for two reasons: first, the differences between the two groups are not so much sharp; and second, we can not discard the presence of some outlying curves, especially among girls.

We perform the first part of the growth data classification study with a similar structure to the simulation study. More precisely, we consider $150$ training samples composed of $40$ and $30$ randomly chosen curves of girls and boys, respectively. We pair each training sample with the test sample composed of the remaining $14$ and $9$ curves of girls and boys, respectively. We denote this way of obtaining training and test samples as T1, and we try to classify the curves included in each test sample by using the methods and depths considered in Section \ref{sec:simStudy}, with the same specifications for both. 

For what concerns the cross validation step for KFSD, we divide each initial training sample in 5 groups-unbalanced cross validation test samples with 8 and 6 curves of girls and boys, respectively, and we pair each of them with its natural cross validation training sample.

We report the performances of the 21 depth-based methods and \textit{k}-NN in Table \ref{tab:growthT3}. In Table \ref{tab:percentilesGrowth} we report the percentages of pairs of initial training and test samples for which the cross validation step is required by DTM+KFSD, WAD+KFSD and WMD+KFSD, and we also show the best performing percentiles for the same samples.

\begin{table}[!htb]
\captionsetup{justification=justified,width=0.75\textwidth}
\caption{Growth data and T1. Means and standard deviations \textit{(in parenthesis)} of the misclassification percentages for DTM, WAD and WMD-based methods and \textit{k}-NN.}
\centering
\scalebox{0.75}{
\begin{tabular}{cccccccc}
\hline
\hline
Method/Depth & FMD & HMD & RTD & IDD & MBD & FSD & KFSD\\
\hline
\multirow{2}{*}{DTM} & 15.22 & 10.84 & 19.33 & 19.91 & 17.30 & 19.45 & 12.14\\
& \textit{(8.18)} & \textit{(7.35)} & \textit{(8.78)} & \textit{(8.99)} & \textit{(8.80)} & \textit{(9.08)} & \textit{(7.82)} \\
\hline
\multirow{2}{*}{WAD} & 14.43 & 11.22 & 15.10 & 15.16 & 14.87 & 15.42 & 12.84\\
& \textit{(7.56)} & \textit{(7.41)} & \textit{(8.18)} & \textit{(7.88)} & \textit{(8.11)} & \textit{(8.12)} & \textit{(7.48)} \\
\hline
\multirow{2}{*}{WMD} & 30.41 & 4.96 & 35.36 & 33.16 & 27.59 & 18.03 & 3.45\\
& \textit{(11.31)} & \textit{(4.56)} & \textit{(8.60)} & \textit{(8.79)} & \textit{(10.22)} & \textit{(7.39)} & \textit{(3.57)} \\
\hline
\hline
\multirow{2}{*}{\textit{k}-NN} & \multicolumn{7}{c}{3.86} \\
& \multicolumn{7}{c}{\textit{(3.56)}}\\
\hline
\end{tabular}
\label{tab:growthT3}
}
\end{table}

Additionally, we consider important to study how the classification methods work when the goal is to classify a single curve using the information contained in the rest of the curves. To do this with the growth data, we consider each possible training sample composed of $92$ curves, and classify the curve not included in the training set. We denote this way of obtaining training and test samples as T2, and we implement a cross validation step for DTM+KFSD, WAD+KFSD and WMD+KFSD also for T2. We report the performances of the classification methods under T2 in Tables \ref{tab:growthNoTrain} and \ref{tab:percentilesGrowth}.

\begin{table}[!htb]
\captionsetup{justification=justified,width=0.75\textwidth}
\caption{Growth data and T2. Number of misclassified curves with DTM, WAD and WMD-based methods and \textit{k}-NN.}
\centering
\scalebox{0.75}{
\begin{tabular}{cccccccc}
\hline
\hline
Method/Depth & FMD & HMD & RTD & IDD & MBD & FSD & KFSD\\
\hline
DTM & 15 & 9 & 17 & 19 & 15 & 18 & 11\\
\hline
WAD & 12 & 10 & 13 & 13 & 13 & 13 & 11\\
\hline
WMD & 28 & 3 & 32 & 30 & 24 & 16 & 2\\
\hline
\hline
\textit{k}-NN & \multicolumn{7}{c}{3}\\
\hline
\end{tabular}}
\label{tab:growthNoTrain}
\end{table}

\begin{table}[!htb]
\captionsetup{justification=justified,width=0.75\textwidth}
\caption{Growth data and DTM+KFSD, WAD+KFSD and WMD+ KFSD. Percentages of initial T1-type and T2-type training and test samples for which cross validation is required and best performing percentiles for the same samples.}
\centering
\scalebox{0.75}{
\begin{tabular}{c|c|c|c|c|c|c|c}
\hline
\hline
T & Method & Percentage & Percentiles & T & Method & Percentage & Percentiles\\
\hline
\multirow{3}{*}{T1} & DTM & 72.00 & 33rd & \multirow{3}{*}{T2} & DTM & 3.23 & 25th\\ 
& WAD & 36.67 & 50th & & WAD & 1.08 & 50th, 66th\\ 
& WMD & 91.33 & 15th & & WMD & 10.75 & 15th\\
\hline
\hline
\end{tabular}}
\label{tab:percentilesGrowth}
\end{table}

The results in Tables \ref{tab:growthT3}-\ref{tab:percentilesGrowth} show that WMD+KFSD is the only method able to outperform \textit{k}-NN, an occurrence that has been already observed with curves generated from CGP4. Indeed, under T1, WMD+KFSD outperforms \textit{k}-NN in terms of means of the misclassification percentages (3.45\% against 3.86\%), whereas the third best method is WMD+HMD (4.96\%); something similar happens under T2: WMD+KFSD misclassifies 2 curves, and it is still the best method, followed by \textit{k}-NN and WMD+HMD, which misclassify 3 curves. If we convert these T2 performances in percentages, i.e., $\left(\frac{\# \mbox{misclassified curves}}{\mbox{sample size}}\times 100 \right)$, we observe that, moving from T1 to T2, there is a slight but systematic improvement: WMD+KFSD, 3.45\% $\rightarrow$ 2.16\%; \textit{k-NN}, 3.86\% $\rightarrow$ 3.23\%; WMD+HMD, 4.96\% $\rightarrow$ 3.23\%. This pattern is due to the greater size of the training samples under T2, but it does not cause significant changes in the performances-based ordering of the methods.

For what concerns the cross validation step of the KFSD-based classification methods, most of the remarks made for the simulated data also hold for this dataset. Moreover, it is clear that the implementation of the cross validation step is a key issue under T1, whereas it becomes much less important under T2. Finally, under both T1 and T2, the best performing percentile for the best method, i.e., WMD+KFSD, is the 15th percentile, which means that classification of growth curves requires a particularly local approach. 

\begin{figure}[!htb]
\centering
\includegraphics[scale=0.5]{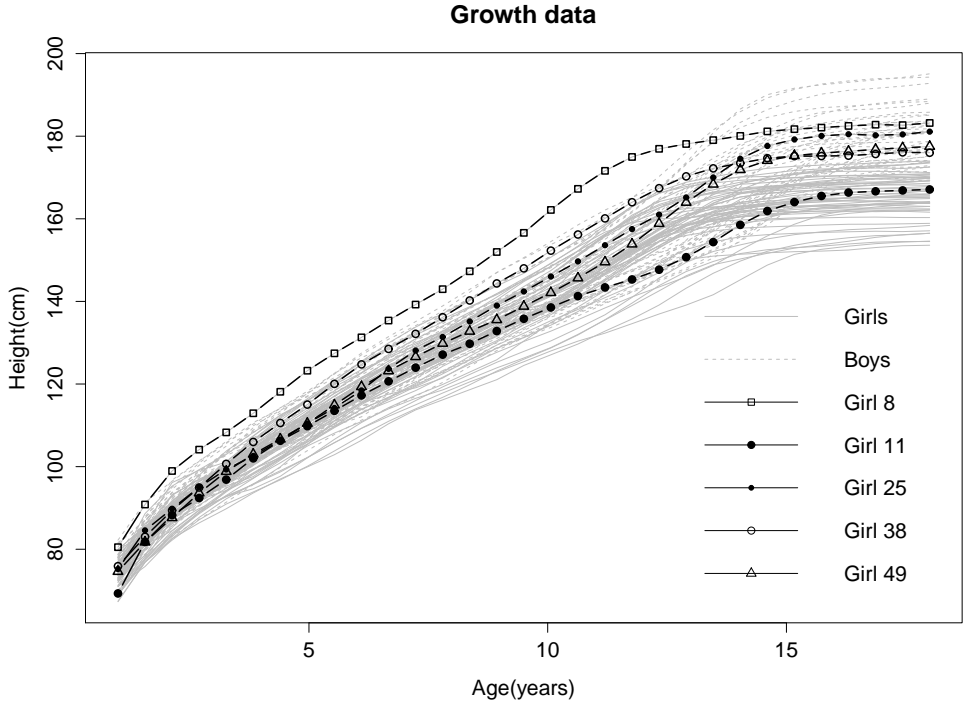}
\captionsetup{width=0.8\textwidth}
\caption{Growth curves: highlighting some interesting curves for the classification problem.}
\label{fig:growthMisCur}
\end{figure}

Even though here we do not show the results obtained with the 7 different percentiles, we would like to report that, when we combine WMD with KFSD and use a fixed percentile, higher percentiles make the performances of the classification method worse. For example, using WMD and KFSD with the 15th and the 25th percentile, we observe means equal to 3.68\% and 4.70\%, respectively, under T1, whereas under T2 the methods misclassify 3 and 4 curves, respectively. Given the results under T2, let us look at the misclassified curves by these two methods and \textit{k}-NN: using the 15th percentile, the method misclassifies girls with labels 11, 25 and 49; using the 25th percentile, the method misclassifies girls with labels 8, 25, 49 and 38; \textit{k}-NN misclassifies girls with labels 8, 25 and 49 (see Figure \ref{fig:growthMisCur}). 

Therefore, the differences between WMD+KFSD used with the 15th percentile and \textit{k}-NN lie in girls 8 and 11. Observing these curves, we can appreciate that with a local spatial approach it is possible to classify correctly a female height having apparently an outlying behavior (Girl 8), however at the price of misclassifying a more central female height (Girl 11); on the contrary, \textit{k}-NN makes the opposite, and its behavior is more similar to the behavior of WMD+KFSD with the 25th percentile, which misclassifies the same curves as \textit{k}-NN, in addition to the girl with label 38. Thanks to the cross validation step, which allows a non-fixed percentile, WMD+KFSD takes advantage of the differences between the use of the 15th and the 25th percentile, and it succeeds in misclassifying only girls with labels 25 and 49.

\subsection{PHONEME DATA}
\label{subsec:phoneme}
\begin{figure}[!htb]
\centering
\includegraphics[scale=0.5]{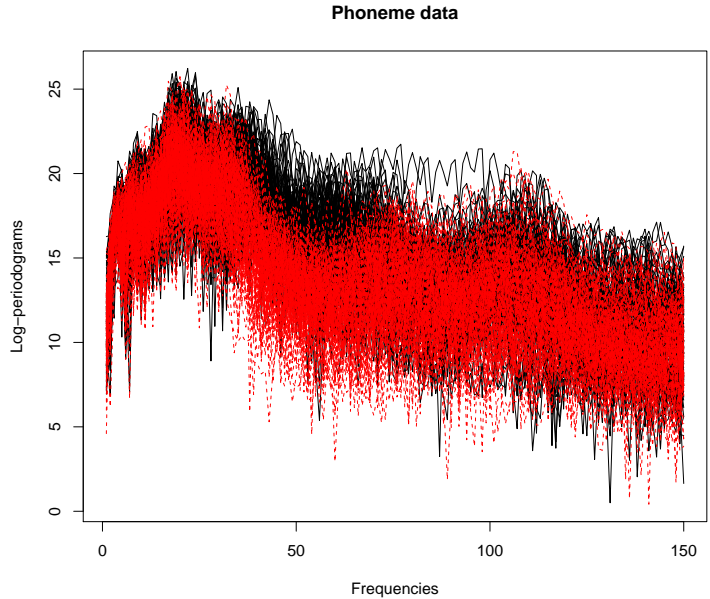}
\captionsetup{width=0.8\textwidth}
\caption{Phoneme data: log-periodograms of 100 AA curves (solid) and 100 AO curves (dashed).}
\label{fig:phoneme}
\end{figure}

The second real dataset that we consider consists in log-periodograms of length $150$ corresponding to recordings of speakers pronouncing the phonemes ``aa'' or ``ao''. More precisely, the dataset contains $400$ recordings of the phoneme ``aa'' and 400 recordings of the phoneme ``ao''. Since we are considering a large number of methods, we perform the study using $100$ randomly chosen recordings of the phoneme ``aa'' (from now on, AA curves) and 100 randomly chosen recordings of the phoneme ``ao'' (from now on, AO curves). Figure \ref{fig:phoneme} shows the curves. For more
details about this dataset, see \cite{FerVie2006}.

Observing Figure \ref{fig:phoneme}, we can appreciate similar features to the ones highlighted for the growth curves: first, we can not discard the presence of some outlying curves in both groups; and second, the differences between the two groups are not so much sharp. Indeed, this second feature seems exaggerated in the second part of the data (frequencies from 76 to 150), and the discriminant information seems to lie especially in the first part of them (frequencies from 1 to 75). This hypothesis has been confirmed by a preliminary classification analysis in which we have observed that in general any method improves its performances when only the first half of the curves is used. Then, we perform the phoneme data classification study using the first 75 frequencies, and with a structure that is similar to the one used for the growth data, that is, we classify curves included in test samples that we obtain by means of both T1 and T2.

To perform the T1 part of the study, we consider 100 training samples composed of 75 randomly chosen AA curves and 75 randomly chosen AO curves. Each training sample is paired with the test sample composed of the remaining 25 AA curves and 25 AO curves (i.e., the allocation ``150 training curves, 50 test curves'' defines T1), and we try to classify the curves included in each test sample by using the same methods and depths as in Section \ref{sec:simStudy}, with the same specifications for both. 

For what concerns the cross validation step for KFSD, it is similar to the one implemented for the simulated data, but in this case we divide each initial training sample in 5 groups-balanced cross validation test samples of size 30 and we pair each of them with its natural cross validation training sample of size 120.

We report the performances of the 21 depth-based methods and \textit{k}-NN in Table \ref{tab:phoneme}, whereas Table \ref{tab:percentilesPhoneme} is the analogous of Table \ref{tab:percentilesGrowth} for the phoneme data.

\begin{table}[!htb]
\captionsetup{justification=justified,width=0.75\textwidth}
\caption{Phoneme data and T1. Means and standard deviations \textit{(in parenthesis)} of the misclassification percentages for DTM, WAD and WMD-based methods and \textit{k}-NN.}
\centering
\scalebox{0.75}{
\begin{tabular}{cccccccc}
\hline
\hline
Method/Depth & FMD & HMD & RTD & IDD & MBD & FSD & KFSD\\
\hline
\multirow{2}{*}{DTM} & 21.56 & 23.16 & 22.68 & 22.70 & 21.84 & 23.00 & 23.08\\
& \textit{(5.07)} & \textit{(5.64)} & \textit{(5.47)} & \textit{(5.37)} &\textit{(5.18)} & \textit{(5.56)} & \textit{(5.60)}\\
\hline
\multirow{2}{*}{WAD} & 23.12 & 23.74 & 23.88 & 23.84 & 23.54 & 23.64 & 23.36\\
& \textit{(5.49)} & \textit{(5.82)} & \textit{(5.78)} & \textit{(5.73)} & \textit{(5.60)} & \textit{(5.82)} & \textit{(5.78)}\\
\hline
\multirow{2}{*}{WMD} & 21.42 & 24.76 & 26.18 & 25.62 & 20.54 & 20.62  & 19.30\\
& \textit{(4.56)} & \textit{(5.50)} & \textit{(5.79)} &  \textit{(6.06)} & \textit{(4.57)} & \textit{(4.86)} & \textit{(4.66)}\\
\hline
\hline
\multirow{2}{*}{\textit{k}-NN} & \multicolumn{7}{c}{22.14} \\
& \multicolumn{7}{c}{\textit{(5.01)}}\\
\hline
\end{tabular}
\label{tab:phoneme}
}
\end{table}

To perform the T2 part of the study, we consider all the possible $200$ training samples composed of $199$ phonemes, each one jointly with its corresponding test sample composed of the remaining curve. As for the growth data, we implement a cross validation step for DTM+KFSD, WAD+KFSD and WMD+KFSD under T2. We report the performances of the classification methods under T2 in Tables \ref{tab:phonemeNoTrain} and \ref{tab:percentilesPhoneme}.

\begin{table}[!htb]
\captionsetup{justification=justified,width=0.75\textwidth}
\caption{Phoneme data and T2. Number of misclassified curves with DTM, WAD and WMD-based methods and \textit{k}-NN-}
\centering
\scalebox{0.75}{
\begin{tabular}{cccccccc}
\hline
\hline
Method/Depth & FMD & HMD & RTD & IDD & MBD & FSD & KFSD\\
\hline
DTM & 43 & 46 & 45 & 46 & 44 & 45 & 46\\
\hline
WAD & 46 & 50 & 51 & 52 & 46 & 49 & 46\\
\hline
WMD & 43 & 51 & 50 & 51 & 39 & 39 & 37\\
\hline
\hline
\textit{k}-NN & \multicolumn{7}{c}{45}\\
\hline
\end{tabular}}
\label{tab:phonemeNoTrain}
\end{table}

\begin{table}[!htb]
\captionsetup{justification=justified,width=0.75\textwidth}
\caption{Phoneme data and DTM+KFSD, WAD+KFSD and WMD+ KFSD. Percentages of initial T1-type and T2-type training and test samples for which cross validation is required and best performing percentiles for the same samples.}
\centering
\scalebox{0.75}{
\begin{tabular}{c|c|c|c|c|c|c|c}
\hline
\hline
T & Method & Percentage & Percentiles & T & Method & Percentage & Percentiles\\
\hline
\multirow{3}{*}{T1} & DTM & 10.00 & 25th & \multirow{3}{*}{T2} & DTM & 0.00 & 15th, 25th, 33rd, 50th, 66th, 75th, 85th\\
& WAD & 33.00 & 15th & & WAD & 1.00 & 15th, 25th, 33rd, 50th, 66th\\
& WMD & 54.00 & 15th & & WMD & 1.50 & 15th, 25th, 33rd, 50th, 66th, 75th\\
\hline
\hline
\end{tabular}}
\label{tab:percentilesPhoneme}
\end{table}

The results in Tables \ref{tab:phoneme}-\ref{tab:percentilesPhoneme} show especially two facts: first, classification of phoneme data is a hard problem, and effectively the number of misclassified curves is considerable with any method and under both T1 and T2; second, WMD+KFSD is the best classification method. Indeed, under T1, WMD+KFSD is the method with the best performance in terms of mean of the misclassification percentages (19.30\%), and it outperforms the second best method, WMD+MBD (20.54\%). The third best method is given by another spatial depth-based method, WMD+FSD (20.62\%). Under T2, WMD+KFSD is again the method with the best performance (37 misclassified curves), whereas WMD+MBD and WMD+FSD are the second best methods (39 misclassified curves). Note that the performance of the fourth best method is quite distant (WMD+FMD, 43 misclassified curves), as well as the one of \textit{k}-NN (45 misclassified curves). If we convert the T2 performances of the three best methods in percentages, there is a slight but systematic improvement when moving from T1 to T2: WMD+KFSD, 19.30\% $\rightarrow$ 18.50\%; WMD+MBD, 20.54\% $\rightarrow$ 19.50\%; WMD+FSD, 20.62\% $\rightarrow$ 19.50\%. However, as for growth data, we observe no significant changes in the performances-based ordering of the methods.

Observing Table \ref{tab:percentilesPhoneme} and focusing on the best KFSD-based methods, i.e., WMD+KFSD, we appreciate that under T1 the best performing percentile for WMD+KFSD is the 15th percentile, whereas for T2 the best performing percentiles for WMD+KFSD are all except the 85th percentile. However, even though we do not show the results obtained with the 7 different percentiles, we would like to report that for the phoneme data, unlike for the growth data, when we combine WMD with KFSD and a fixed percentile, even with the worst performing percentile, which is the 85th percentile, WMD+KFSD has performances comparable to the best ones. Indeed, using the 85th percentile, WMD+KFSD still outperforms the second best method of Table \ref{tab:phoneme} (19.90\% against 20.54\% of WMD+MBD under T1), and it misclassifies the same number of curves as the second best methods of Table \ref{tab:phonemeNoTrain} (39 misclassified curves by WMD+MBD and WMD+FSD under T2).

\section{CONCLUSIONS}
\label{sec:conc}

In this paper we have introduced a new functional depth, the kernelized functional spatial depth, that represents a local-oriented and kernel-based version of the functional spatial depth recently proposed by \cite{ChaCha2013}. Originally developed by \cite{Cha1996} and \cite{Ser2002} in the multivariate context, the spatial approach allows FSD and KFSD to study the degree of centrality of curves from a new point of view with respect to the other existing functional depths. The main novelty introduced by KFSD consists in the fact that it addresses the study of functional datasets at a local spatial level, whereas FSD is more appropriate for global spatial analyses. 

As we showed in Section \ref{sec:FSDs}, KFSD and FSD are related: $KFSD(x,P) = FSD(\phi(x) , P_{\phi})$, where $\phi: \mathbb{H} \to \mathbb{F}$ is an embedding map, $\mathbb{F}$ is a feature space and $P_\phi$ is a recoded version of the probability distribution $P$. The embedding map $\phi$ and the feature space $\mathbb{F}$ are implicitly defined through a positive definite kernel, $\kappa: \mathbb{H} \times \mathbb{H} \rightarrow \mathbb{R}$. We think that the above-mentioned relationship is a key feature to analyze the theoretical properties of KFSD in the future. 

Afterwards we focused on supervised functional classification problems, especially in situations in which the differences between groups are not excessively marked and/or the data may contain outliers. We have studied the classification performances of three depth-based methods, DTM, WAD and WMD, and a benchmark procedure such as \textit{k}-NN. The three depth-based methods have been used together with KFSD, FSD and five more existing functional depths. In general, we have observed that a KFSD-based method is always among the best methods in terms of classification capabilities for the considered simulation and real scenarios, and that it outperforms the benchmark procedure \textit{k}-NN. Note that no other depth behaves as well as KFSD, and that WMD+KFSD produces doubtless the most stable and best depth-based classification method: indeed, WMD+KFSD always outperforms WMD+FSD, which is its natural global-oriented competitor, and, more in general, it has always acceptable results, which are often the best ones, as in the case of the considered real datasets.

{
\section*{ACKNOWLEDGMENTS}
\indent The authors would like to thank the associate editor and four anonymous referees for their helpful comments. This research was partially supported by Spanish Ministry of Education and Science grant 2007/04438/001, by Spanish Ministry of Science and Innovation grant 2012/00084/001, and by MCI grant MTM2008-03010.
}

\bibliographystyle{asa}
\bibliography{paper01_20130513}

\end{document}